\documentclass[final,5p,times]{elsarticle}

\usepackage{graphicx}
\usepackage{amssymb}
\usepackage{amsthm}

\usepackage{amsmath}
\usepackage{units}
\usepackage{multirow}
\usepackage{float}
\usepackage{rotating}

\biboptions{sort&compress}

\journal{Solid State Nuclear Magnetic Resonance}

\begin{document}

\begin{frontmatter}

\title{First Principles NMR Study of Fluorapatite under Pressure}

\author[SAM]{Barbara Pavan\corref{cor1}}
\ead{pavan1b@cmich.edu}
\cortext[cor1]{Corresponding author. Fax: +1 (989) 774 - 2697}

\author[CNR]{Davide Ceresoli}
\author[SAM,CHM]{Mary M. J. Tecklenburg}
\author[SAM,PHY]{Marco Fornari}

\address[SAM]{Science of Advanced Materials, Central Michigan University, Mt. Pleasant, MI 48859, USA}
\address[CNR]{CNR, Institute of Molecular Science and Technology (ISTM), via Golgi 19, 20133 Milan, Italy}
\address[CHM]{Department of Chemistry, Central Michigan University, Mt. Pleasant, MI 48859, USA }
\address[PHY]{Department of Physics, Central Michigan University, Mt. Pleasant, MI 48859, USA }

\begin{abstract}
NMR is the technique of election to probe the local properties of materials. Herein we present the results of density functional theory (DFT) \textit{ab initio} calculations of the NMR parameters for fluorapatite (FAp), a calcium orthophosphate mineral belonging to the apatite family, by using the GIPAW method [Pickard and Mauri, 2001]. 
Understanding the local effects of pressure on apatites is particularly relevant because of their important role in many solid state and biomedical applications. Apatites are open structures, which can undergo complex anisotropic deformations, and the response of NMR can elucidate the microscopic changes induced by an applied pressure. 
The computed NMR parameters proved to be in good agreement with the available experimental data.  The structural evaluation of the material behavior under hydrostatic pressure (from --5 to +100 kbar) indicated a shrinkage of the diameter of the apatitic channel, and a strong correlation between NMR shielding and pressure, proving the sensitivity of this technique to even small changes in the chemical environment around the nuclei. This theoretical approach allows the exploration of all the different nuclei composing the material, thus providing a very useful guidance in the interpretation of experimental results, particularly valuable for the more challenging nuclei such as $^{43}$Ca and $^{17}$O.
\end{abstract}


\end{frontmatter}

\section{Introduction}\label{Intro}

Calcium orthophosphates, and in particular apatites (general formula Ca$_{5}$(PO$_4$)$_3$X (X=F,Cl,OH)), are extensively studied because of their significance in many fields, from geology and materials science to medicine. Apatites, with various degrees of substitutions and defects, are commonly found in the Earth's crust and are considered one of the most abundant sources of phosphorous in the marine environment, acting as phosphorous ``sinks" \cite{Tribble1995}. Apatites are also attracting a great interest in the field of biomaterials due to their intrinsic biocompatibility and bioactivity.
Calcium phosphates, such as synthetic hydroxyapatite (HAp), fluorapatite (FAp), $\alpha$ and $\beta$-tricalcium phosphate ($\alpha$-TCP, $\beta$-TCP), displayed intrinsic osteoinductive properties, namely the capacity of promoting bone formation without presence of oesteogenic factors, as well as osteoconductive properties (the capacity of supporting and guiding the growth of newly forming bone). Apatite bioceramics are used in bone repair and augmentation, as bone cements, as coatings for bio-inert prostheses,  and for bone scaffolding \cite{Dorozhkin2010,LeGeros2008}.  
 The study of natural apatites is complicated by the fact that the apatite crystal structure can readily accept substitution(s) from a large variety of ions. For example, fluoride ions are found to readily substitute in human enamel, due to their presence in drinking water as a result of soil erosion or added on purpose for caries prevention policies \cite{RosinGrget2001}. This capability of accepting various substitutions and incorporating even bulky ions, such as Cd(II) and Cu(II), as well as U(IV), is making this class of materials also attractive for nuclear waste management and water remediation \cite{Zheng2007, Wellman2008}.
 
Fluorapatite (FAp) is one of the most well-characterized members of the apatite family, and in materials science it has found interesting applications  as a host for solid state diode lasers when doped with rare-earth elements, such as praseodymium or ytterbium \cite{Sardar2004}. It can be also used as a phosphor for fluorescent lamps or plasma display panels, exploiting its photoluminescent properties when doped with Sr, or rare earth elements \cite{Liang2007,Rakovan2000}.

Good knowledge of the local structural properties and the response of FAp to external stimuli is of paramount importance to gain understanding of this material and, due to its structural similarity, the study of FAp can be considered as a reference for the other members of the apatite family.  Solubility, crystallization, dissolution, and ion exchange have been intensively studied \cite{Wang2008,Bengtsson2009}, but relatively few studies explored the behavior of FAp under pressure and they were mainly focused on the bulk properties of the material, studied by using vibrational spectroscopic techniques and X-ray diffraction\cite{Comodi2001,Comodi2001_2,Matsukage2004,Williams1996}.

In the present work we aim to explore the effects of pressure at the local level, by first principles calculations of the NMR parameters response. NMR is the technique of election for studying the local structure of materials since it is very sensitive to changes in the atomic environment. A deformation occurring around a nucleus will in fact affect its shielding, thus providing information on the deformation itself. In particular, our theoretical approach allows us to obtain unambiguously all the components of the shielding tensor and therefore to have a clear three dimensional picture of the effects of pressure on each nucleus in the material.

The \textit{ab initio} modeling of apatites is demanding from a computational point of view and was performed only in few cases in the past \cite{Peeters1997, Louis-Achille1998,Leeuw2002,Leeuw2002-2,Calderin2003,Sahai2005, Tamm2006, Astala2006,Peroos2006, Rulis2007, Zahn2008, Matsos2009}. The published works have been mainly focused on structural and electronic features: a first principles study of fluorapatite and hydroxyapatite has been recently published by Rulis et al. (Ref.~\cite{Rulis2007}) where the authors computed electronic structure, charge distribution and X-ray absorption spectra on a supercell slab along  the (001) surface. Some theoretical NMR studies are available in the literature, but they are mainly focused on hydroxyapatite \cite{Chappell2008,Gervais2008,Pourpoint2007}. First principles calculations of NMR parameters have been available for molecules and clusters since the `70s \cite{GIAO,IGLO,IGAIM,CSGT, Schindler82} while their application to periodic/crystalline systems was implemented at a later time~\cite{Mauri96} due to the difficulty of including a macroscopic magnetic field, requiring a non-periodic vector potential. In 2001, Pickard and Mauri implemented the Gauge Including Projector Augmented Wave (GIPAW) method~\cite{Pickard2001} in the plane-wave pseudopotential (PWPP) framework, based on linear-response, aimed at calculations of periodic systems. In the GIPAW method, the key to an accurate evaluation of NMR chemical shifts is the reconstruction of the all-electron wavefunction from a pseudopotential calculation, by a modified PAW transformation, in order to describe the nodal structure of the wavefunction close to the nuclei. Hence the GIPAW method retains the accuracy of all-electron calculations at the much lower cost of pseudopotential calculations, i.e. explicit treatment of valence electrons.

In the present work, the structural properties of FAp will be evaluated when hydrostatic pressure is applied in the range from --5 to +100 kbar, and particular focus will be dedicated to the response of the NMR shielding. For the nuclei with spin $I>\nicefrac{1}{2}$ the quadrupolar coupling constant (C$_q$) and the asymmetry parameter ($\eta_q$) will also be provided. 
 This paper is organized as follows: in sections \ref{methods} and \ref{structure} we discuss the methodology and the structural data, in sections \ref{results1} and \ref{results2} we present and discuss our NMR calculations results, and conclusions are in section \ref{conclusions}.
 
\section{Computational methods}\label{methods}

The calculations of fluorapatite (FAp) have been performed by using \textit{ab initio} density functional theory (DFT) \cite{Parr1989} as implemented in the Quantum-ESPRESSO package \cite{Giannozzi2009}, with the PBE exchange-correlation functional \cite{PBE1996}.
Well converged plane wave basis sets were employed for all the presented calculations, using norm-conserving GIPAW pseudopotentials with a kinetic energy cut-off chosen at $E_{cut}= 80$ Ry.  The convergence criteria for the total energy was set in all cases at 10$^{-8}$ Ry. The \textbf{k}-points sampling in the Brillouin zone (BZ) used a Monkhorst-Pack grid with a converged mesh of $2\times2\times2$. 

The FAp structure used for comparing the computed NMR parameters with the experimental data was constructed using the experimental lattice parameters (Ref.~\cite{Sudarsan1972}) and fully relaxed internal degrees of freedoms.
The structures used for the evaluation of the NMR response under an applied pressure were obtained by imposing a target pressure (from --5 to 100$\pm1$ kbar), and relaxing not only the atomic positions, but also the lattice parameters (Wentzcovitch damped dynamic algorithm, Ref.~\cite{Wentzcovitch1991_2}). The only constraint imposed on these relaxations was the conservation of symmetry, under the assumption that no phase transition would occur at small enough pressures. This variable cell relaxation approach well simulates hydrostatic pressure for bulk systems.
 
NMR shieldings critically depend on the electronic wavefunctions in proximity of the nucleus. However, those are explicitly neglected in the pseudopotential approach, where the electronic wavefunctions are described by smooth functions and thus their correct nodal structure is lost. In the present work, specially designed norm-conserving GIPAW pseudopotentials were used\footnote{We used standard Martins-Troullier pseudization. For oxygen and fluorine, the local channel is $p$ and cutoff radii are 1.3 and 1.4 atomic units, respectively. For phosphorous, the local channel is $d$ and cutoff radius is 1.9. For calcium, the local potential is derived from the $d$ channel, and cutoff radii are 1.45, 2.00, 1.45 for the $s$, $p$, and $d$ channel respectively.}  which allowed the calculation of the NMR shieldings. 
We also used a modified calcium pseudopotential, constructed with a rigid shift of the 3d orbitals by +3.2 eV, for the calculation of Ca and O NMR shieldings. The choice of using this pseudopotential was based on the observation by Profeta \textit{et al.} \cite{Profeta2004} that in calcium oxides and calcium aluminosilicates the Ca PBE functional tends to overestimate the degree of covalency of the Ca--O bonds, thus causing the shielding on both Ca and O to be too small. However, this correction on the Ca pseudopotential had a negative influence on the computed equilibrium volumes and the computed phonon frequencies. Therefore in the present work, the structural relaxations and the NMR shieldings of F and P were obtained by using unit cells computed with the unshifted Ca pseudopotential, while the use of the shifted pseudopotential was limited to the calculation of Ca and O NMR parameters, assuming that the underprediction of the NMR shielding was a problem limited to Ca--O interactions, and that this effect was not relevant when calcium was present in different atomic environments. 

The NMR absolute chemical shielding tensor ($\overleftrightarrow{\sigma}$) is defined as $\mathbf{B^{ind}}=-\overleftrightarrow{\sigma}(\mathbf{r})\cdot \mathbf{B^{ext}}$, where $\mathbf{B^{ext}}$ is the externally applied uniform magnetic field. $\mathbf{B^{ext}}$ generates a local current at the nucleus, which in turn induces a local magnetic field $\mathbf{B^{ind}}$ so that the total magnetic field experienced by the nucleus is $\mathbf{B}=\mathbf{B^{ext}}+\mathbf{B^{ind}}$. 

Several different conventions for notation are in current use: in the present work the ``Haeberlen notation'' is used \cite{Harris2008}.
We report the isotropic NMR shielding ($\sigma_{iso}$) as the average of the principal components of the shielding tensor ($\overleftrightarrow{\sigma}$), expressed in a suitable set of axes $X$, $Y$, and $Z$:
\begin{equation*} 
\sigma_{iso}=\dfrac{1}{3}(\sigma_{XX}+\sigma_{YY}+\sigma_{ZZ}),
\end{equation*}
where the three principal components are related as follows:
\begin{equation*} 
|\sigma_{ZZ}-\sigma_{iso}|\geq |\sigma_{XX}-\sigma_{iso}|\geq |\sigma_{YY}-\sigma_{iso}|.
\end{equation*} 

To fully describe the local symmetry around the nucleus, the shielding anisotropy ($\zeta$) is also reported as:
\begin{equation*} 
\zeta= \sigma_{ZZ}-\sigma_{iso};
\end{equation*}
as well as the asymmetry parameter ($\eta$) as:
\begin{equation*} 
\eta=\dfrac{\sigma_{YY}-\sigma_{XX}}{\zeta}.
\end{equation*} 



Finally, for the nuclei with $I>\nicefrac{1}{2}$ ($^{17}$O and $^{43}$Ca), the quadrupolar interactions are described by their quadrupolar coupling constant ($C_q$) and the asymmetry parameter ($\eta_q)$, defined as follows:

\begin{equation*} 
C_q=\dfrac{eV_{ZZ}Q}{h}
\end{equation*}
\begin{equation*} 
\eta_q=\dfrac{V_{XX}-V_{YY}}{V_{ZZ}}
\end{equation*}

\noindent where $e$ is the elemental charge, $h$ is the Planck constant, and $Q$ is the nuclear quadrupole moment ($Q$($^{17}$O)$=-2.56\times10^{-30}$ m$^2$, $Q$($^{43}$Ca)$=-4.90\times10^{-30}$ m$^2$; Ref.~\cite{Raghavan1989}). $V_{xx}$, $V_{yy}$, and $V_{zz}$ are the diagonal elements of the electric field gradient (EFG) tensor in the principal axis system (PAS), with $\mid V_{ZZ}\mid \geq\mid V_{YY}\mid \geq\mid V_{XX}\mid$.

Most of the experimental NMR data are reported normalized in ppm as chemical shifts ($\delta_{iso}$),  with respect to a reference compound, where the chemical shift of the sample is obtained as:

\begin{equation*} 
\delta_{sample}=\dfrac{\nu_{sample}-\nu_{ref}}{\nu_{ref}}(\times 10^6)
\end{equation*}

\noindent where $\nu_{sample}$ and $\nu_{ref}$ are the resonance frequencies of the sample and the reference compound, respectively. 

The chemical shift and the chemical shielding are related by the following:

\begin{equation*} 
\delta_{sample}=\sigma_{ref}-\sigma_{sample}
\end{equation*}
To readily compare the computed data with the corresponding experimental values, it is necessary to calculate their chemical shift by choosing an appropriate reference material. Experimentally, most of the reference compounds for solid state NMR are molecules dissolved in water or an organic solvent, which are not well suited for calculations with our methodology. Our approach for overcoming this obstacle was therefore to choose solid reference compounds (\textit{secondary reference}), with known experimental chemical shifts with respect to the standard liquid reference compounds (\textit{primary reference}) recommended by IUPAC \cite{Harris2002}. In Table~\ref{shielding-ref}  the solid reference compounds  used in the present study are reported, along with  their experimental chemical shifts ($\delta_{iso}^{ref}$) and their computed isotropic shielding values ($\sigma_{iso}^{ref}$). 
When more than one value for the experimental chemical shift was available, a range of computed NMR shifts is reported.  

The computed NMR chemical shifts ($\delta_{iso}$), for comparison with experiments, described in the present work are calculated as follows:
\begin{equation*}
\delta_{iso}= \delta_{iso}' +\delta_{iso}^{ref}= (\sigma_{iso}^{ref}-\sigma_{iso}) +\delta_{iso}^{ref}
\end{equation*} 

\noindent where $\delta_{iso}'$ is the computed chemical shift with respect to the secondary reference material, while $\delta_{iso}$ is the chemical shift with respect to the primary reference material, which by definition has a $\delta=0$ ppm. The $\sigma_{iso}^{ref}$ is the computed isotropic shielding for the secondary reference compound, while $\sigma_{iso}$ is the isotropic shielding computed for the nucleus of interest. 

\begin{table}[htbp]
\caption{Primary reference materials ($\delta=0$ ppm) and secondary reference materials with their experimental isotropic chemical shift ($\delta_{iso}^{ref}$ with respect to the primary references \cite{Mackenzie2002}), and their computed NMR isotropic shielding ($\sigma_{iso}^{ref}$).} 
\begin{center}
\resizebox{8.75cm}{!}{
\begin{tabular}{c c  ccc}
\hline
        &Primary & \multicolumn{3}{c}{Secondary reference}\\
       &reference& & $\delta_{iso}^{ref}$& $\sigma_{iso}^{ref}$\\
Nucleus&$\delta_{iso}=0$ (ppm) & Material& (ppm) &(ppm)\\
\hline 
\multirow{2}{*}{$^{43}$Ca} & \multirow{2}{*}{CaCl$_{2(aq.)}$ (1.0M)}& \multirow{2}{*}{CaO$_{(s)}$}& \multirow{2}{*}{136}& \multirow{2}{*}{1032.79}\\
 & & & & \\
 \hline
\multirow{2}{*}{$^{19}$F}& \multirow{2}{*}{CFCl$_{3(l)}$}&\multirow{2}{*}{CaF$_{2(s)}$}&$-107$ &\multirow{2}{*}{219.90} \\
	&	&	&$-104.8$ &	\\
\hline
\multirow{2}{*}{$^{31}$P}& \multirow{2}{*}{H$_3$PO$_{4(aq.)}$ 85\% }&\multirow{2}{*}{BPO$_4$} & $-29.5$ &\multirow{2}{*}{317.82} \\
	&	&	& $-31.2$ &	\\
\hline
\multirow{2}{*}{$^{17}$O }&\multirow{2}{*}{ D$_2$O$_{(l)}$ }& \multirow{2}{*}{quartz} & \multirow{2}{*}{$38\pm2$} & \multirow{2}{*}{211.84}\\
 & & & & \\
\hline
\end{tabular}}
\end{center}
\label{shielding-ref}
\end{table}

The proposed methodology of referencing the computed NMR shielding values allows for a direct comparison with most of the reported experimental data. Moreover it allows the referencing of our calculated values  to systems which are computationally more similar to the ones studied in the present work, avoiding the complications inherent in simulating molecules in liquids. 

\section{Structural data} \label{structure}

Fluorapatite has a hexagonal unit cell with $P6_3/m$ space group (n.~176), containing two formula units (42 atoms) per unit cell. The structure used in the present work is based on the crystallographic data of Sudarsan \textit{et al.} (Ref.~\cite {Sudarsan1972}).  The nonequivalent atoms after relaxation of their positions are presented in Table \ref{FAp-structure}. The relaxed structure is very close to the experimental: the largest difference was  of $-0.12$~\AA~ on the $y$ coordinate of Ca(2), while all the remaining atomic coordinates were within $\pm0.02$~\AA~ of the experimental values. 

\begin{table} [htbp] 
\caption{Fluorapatite unit cell parameters and nonequivalent atoms obtained after relaxation of their atomic positions. Space group $P6_3/m$ (n.~176).} 
\begin{center}
\resizebox{8.75cm}{!}{
\begin{tabular}{lcccccc}
\hline 
Lattice   \\
Parameters &Atom &Wyckoff position & $x$ & $y$ & $z$\\
\hline 
a=b=9.363\AA 	& Ca(1) 	& $f$ 	&1/3	&2/3	& 0.0006\\
c=6.878\AA	       &Ca(2)	& $h$ 	& 0.2417&-0.0054	& 1/4	\\
$\alpha$=$\beta$=90$^\circ$& P 		& $h$ 	&0.3978	&0.3686	&1/4	\\
$\gamma$=120$^\circ$ & O(1)	& $h$	&0.3254	&0.4858	& 1/4	\\
& O(2)	& $h$	&0.5905	&0.4680	&1/4	\\
& O(3)	& $i$	&0.3400	&0.2546	& 0.0684\\
& F		& $a$	& 0		&0		&1/4	\\
\hline 

\end{tabular}}
\end{center}
\label{FAp-structure}
\end{table}

\begin{figure}[H]
\begin{center}
\includegraphics{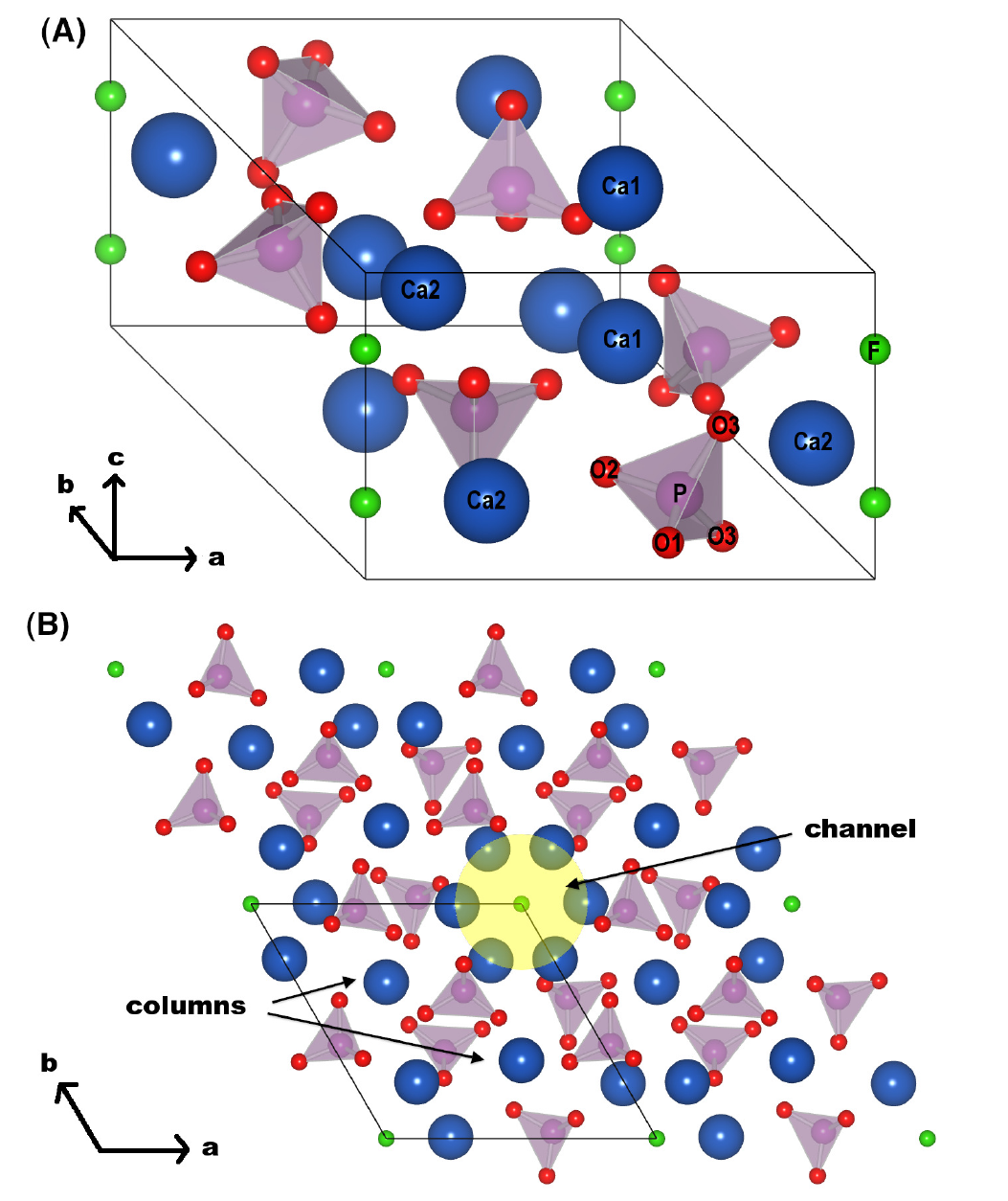}
\caption{(A) Fluorapatite unit cell and (B) four adjacent fluorapatite unit cells depicted along the c-axis. The channel (shadowed region) and the columnar calciums are indicated. (Ca = blue, F = green, P = purple, O = red; the images were obtained using VESTA \cite{Momma2008})}
\label{fig1}
\end{center}
\end{figure}

The fluorapatite unit cell, and four adjacent FAp unit cells seen along the c-axis are presented in Fig.~\ref{fig1}(A) and \ref{fig1}(B), respectively.
Two distinct Ca sites are present in the unit cell: Ca(1) and Ca(2).  Four Ca(1) occupy two vertical columns parallel to the c-axis, and are usually referred to as ``columnar". The remaining six Ca(2) can be divided into two groups of three calcium atoms, disposed to form two triangles lying on two planes perpendicular to the c-axis. The two triangles formed by the Ca(2) are respectively rotated by 60$^\circ$ and they delimit the apatitic channel where the fluorine ions are located. The presence of this channel is a feature common to the minerals with apatitic structure.
Three nonequivalent oxygens are also present, and they are arranged to form six slightly distorted tetrahedra around the six equivalent phosphorous atoms. In particular, each tetrahedron contains one O(1) and O(2), and two O(3). The oxygen atoms form a network connecting Ca(1) and Ca(2) atoms, which are nine-fold and six-fold coordinated with oxygens, respectively.

\section{Computed NMR and comparison with experiments}\label{results1}

\begin{sidewaystable}
\begin{center}
\caption{Calculated NMR isotropic shieldings ($\sigma_{iso}$), and chemical shifts ($\delta_{iso}$) for fluorapatite, along with experimental data when available.  Computed shielding anisotropy ($\zeta$), asymmetry ($\eta$), and the quadrupolar parameters $C_q$ and $\eta_q$ are also provided.}

\begin{tabular}{ccccccccc}
\hline 
 & $\sigma_{iso}$  & \multicolumn{2}{c}{$\delta_{iso}$ (ppm)} & $\zeta$  & \multicolumn{2}{c}{$\eta$} & $C_{q}$ & \multirow{2}{*}{$\eta_{q}$}\\
\cline{3-4} \cline{6-7} 
 & (ppm) & Calc. & Expt. \cite{Braun1995,Dupree1997,Mason2009} & (ppm) & Calc.  & Expt. \cite{Braun1995}  & (MHz) & \\
\hline
Ca(1) & 1170.72 & -1.93 & \multirow{2}{*}{ -14}  & 33.25 & 0.00 &  & 2.90 & 0.00\\
Ca(2) & 1170.53 & -1.74 &  & -26.05 & 0.67 &  & -2.31 & 0.89\\
O(1) & 122.64 & 127.20 &  & 57.80 & 0.04 &  & -5.18 & 0.15\\
O(2) & 123.81 & 126.03 &  & 28.27 & 0.42 &  & -5.11 & 0.05\\
O(3) & 132.75 & 117.09 &  & 37.66 & 0.11 &  & -5.07 & 0.18\\
P & 285.50 & 1.12, 2.82 & 2.3, 2.6 & -8.68 & 0.64 & 0.4--0.5 & n/a & n/a\\
F & 206.48 & -(94.29--92.09) & -100.6, -99.8 & -84.32 & 0.00 & 0.2--0.7 & n/a & n/a\\
\hline
\end{tabular}

\label{FAp-NMR}
\end{center}
\end{sidewaystable}

The crystal structure, as described in Table~\ref{FAp-structure}, was used for the NMR parameters calculations, and the results are reported in Table~\ref{FAp-NMR}. Our calculated NMR chemical shifts are in good agreement with the experimental values available, especially for $^{31}$P where it is within 1 ppm. For the calcium and fluorine values, they are qualitatively similar to the experimental data, within  $\approx$13 and 8 ppm, respectively.  All the nuclei showed a large shielding anisotropy, greater than 25 ppm (absolute value), with the exception of phosphorous atoms where it was about --8 ppm, thus indicating that the shielding around the P has a lower anisotropy due to the higher local symmetry of the phosphate group. 

According to the available literature, the only experimental work on $^{43}$Ca for FAp analyzed a fluoridated hydroxyapatite with a not specified fluorine content (Ref.~\cite{Dupree1997}, w.r.t. CaCl$_{2}$ sat.).  In that work the two distinct calcium sites were not resolved, although some asymmetry in the lineshape was evident. In Table \ref{FAp-NMR} we report this measured $^{43}$Ca chemical shift as the peak maximum, adjusted to a CaCl$_{2}$ (1.0 M) reference solution (+8 ppm). Throughout this paper all the reported $^{43}$Ca chemical shift values are adjusted to this specific reference, which gives a $^{43}$CaO chemical shift at 136 ppm (about 6--8 ppm higher than when referenced to a CaCl$_{2}$ saturated solution). This approach does not totally eliminate the discrepancies in the calcium referencing, since saturated solutions prepared from anhydrous or different hydrates of CaCl$_{2}$ can cause the $^{43}$Ca chemical shift to cover a range of about 7 ppm, between +8.5 and +15.5 ppm. For a detailed description of how the calcium chemical shift is affected by differently prepared CaCl$_{2}$ solutions, consult Gervais \textit{et al.} (Ref.~\cite{Gervais2008}). 
The few additional works regarding $^{43}$Ca are for hydroxyapatite, placing the $^{43}$Ca chemical shifts at 4.5--10 ppm and 17.5--22 ppm for Ca(1) and Ca(2), respectively \cite{Gervais2008,Laurencin2008,Xu2010}. Even considering the possible discrepancies in the referencing, and the unspecified composition of the fluorapatite mineral analyzed in Ref.~\cite{Dupree1997}, these data suggest that when fluorine is present in the apatitic channel it causes an increase in the shielding on the Ca nuclei (more negative chemical shift value), a trend which is well reproduced by our calculations. Moreover, our calculations also showed that the presence of fluorine caused a smaller difference in shielding values between the two distinct calcium sites, when compared to hydroxyapatite.

To the best of our knowledge, no experimental work was done on $^{17}$O for FAp, even though some data are available for hydroxyapatite, with $^{17}$O chemical shifts at 108 and 115 ppm (unassigned) \cite{Wu1997}. A direct comparison of our data with these published values can potentially lead to errors, even though the crystal structures of the two compounds are similar. Nevertheless, we do not expect large differences in the $^{17}$O chemical shift within the apatite class of compounds, since the environment immediately around the oxygen atoms is scarcely modified by different ions filling the apatitic channel. 
Our computed values for $^{17}$O are in fact within approximately 11 ppm of the experimental values for HAp. Only two types of oxygens were detected in the experiment, even if three crystallographically distinct oxygens exist in the apatitic structure.   From our calculations we found that the O(1) and O(2) chemical shifts have very similar values, indicating a similar chemical environment, whereas the chemical shift of O(3) is clearly different, and it is separated from the others by about 9--10 ppm. This observation is in agreement with that found experimentally by Wu \textit{et al.} (Ref.~\cite{Wu1997}) where the two peaks were separated by 7 ppm. If we  approximate that the  $^{17}$O chemical shifts in FAp and HAp are similar, we can assign the more downfield peak found in HAp (115 ppm) to O(1) and O(2), and the more upfield signal (108 ppm) to O(3). 

The quadrupolar coupling constants calculated in the present work for the two crystallographically distinct Ca nuclei had similar values of 2.3 and 2.9 MHz, but opposite sign. The experimentally reported quadrupolar coupling constants for calcium (in HAp)  are of 2.6 MHz for both Ca(1) and Ca(2) \cite{Laurencin2008,Xu2010}. The difference in sign of the $C_q$ between the two distinct calcium nuclei found in our calculations was not observed experimentally (at least in HAp), but it was found in calculations performed by Gervais \textit{et al.} \cite{Gervais2008} on HAp. 
All the computed $C_q$ for oxygens displayed a quite large and negative value of about -5 MHz. If our comparison with HAp holds, we can say that the the absolute value of the oxygens $C_q$ is similar to the ones experimentally determined by Wu \textit{et al.} (4.0 and 4.1 MHz) \cite{Wu1997}.

Although our calculated chemical shifts are in good agreement with the experimentally available values, some of the noted discrepancies can arise mainly from two facts: (1) variability in the preparation/composition of the primary reference compound, and (2) the error in the NMR parameters calculations. The variability in the preparation of the reference compounds becomes of greater influence for more challenging, low abundance and low sensitivity nuclei, such as $^{43}$Ca and $^{17}$O, where also fewer experiments are available to compare with. This error can affect the value of both the primary and secondary reference chemical shifts, thus generating a larger difference from the computed chemical shifts for those nuclei. For more abundant nuclei, where referencing is less challenging and more reproducible, the discrepancy between calculated and experimental values is reduced.

\section{Structural evaluation and NMR calculations under pressure}\label{results2}

The FAp structure as described in Table~\ref{FAp-structure} was fully relaxed without constraining either its lattice parameters, or its symmetry. The relaxed structure so obtained recovered all the original symmetries, while the resulting equilibrium volume was slightly bigger than the values reported in Table \ref{FAp-structure} (a = +1.6\%, c= +0.22\%). The tendency of overestimating the volume is a well known feature of the PBE functional. 

A set of structures were subsequently obtained by imposing a target pressure ranging from --5 to +100 kbar, and the computed energies and volumes for all these structures were used to calculate the bulk modulus of the material ($K_0$), by fitting with the Murnaghan equation of state. The computed bulk modulus was $K_0=94 \pm 2$ GPa, in good agreement with the experimental results reported by Comodi \textit{et al.} (Ref.~\cite{Comodi2001}) of 93 GPa.
The whole set of calculations at finite pressure did not cause major changes in the atomic positions within the crystal structure, nevertheless, when pressure was applied it caused a contraction of the a and b-axis which was more pronounced than the c-axis. This result is also in agreement with experimental measurements of FAp under hydrostatic pressure, obtained using in situ single crystal X-ray diffraction, in the 0--7 GPa pressure range \cite{Comodi2001}. This suggests that the c-axis is stiffer than a and b, possibly due to the arrangements of Ca(1) and F atoms into vertical columns along the c-axis. This columnar arrangement can cause a stronger electrostatic repulsion between the atoms within the columns, whereas the atoms that are staggered, such as Ca(2) and O atoms, can more freely rearrange in response to the applied pressure. 
The computed structures were further analyzed by calculating their pair distribution functions (PDF), using the ISAACS software \cite{Le-Roux:2010}. This allowed us to explore the distribution of the atoms within the unit cell, and how the distances between the atoms changed when pressure was applied. Most of the changes in the atomic distances occurred between Ca(2)--F and Ca--O (Ca(2)--O(3) and Ca(1)--O(2)), suggesting a shrinkage of the apatitic channel and a closer packing of the phosphates, respectively. The P--O and O--O distances were essentially not affected by pressure, thus indicating that the phosphate units are more rigid than their surroundings, and confirming the larger degree of covalency of the P--O bonds. 

\begin{figure}[htbp] 
\begin{center}
\includegraphics{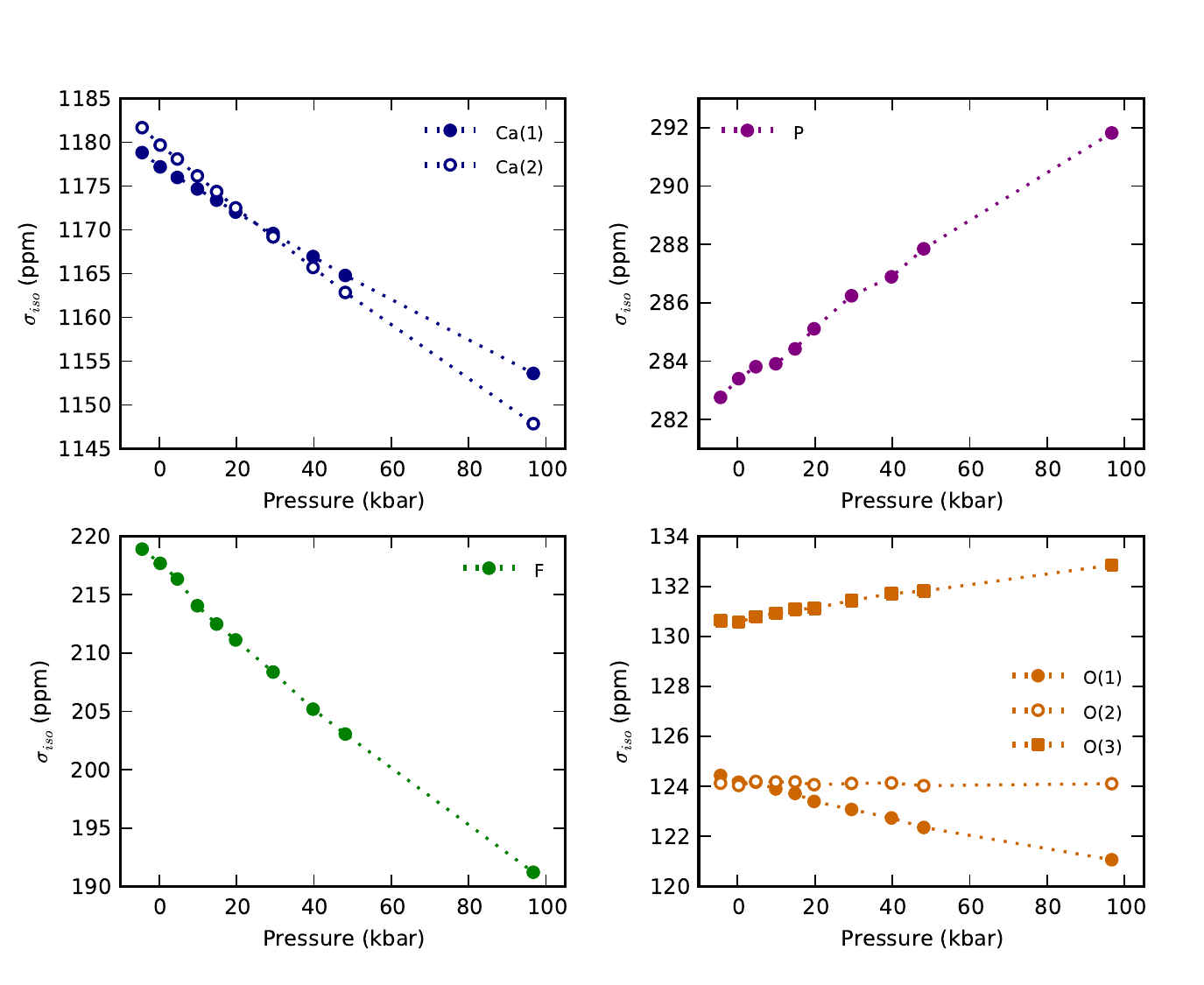}
\caption{The NMR isotropic shielding ($\sigma_{iso}$) for the different nuclei in fluorapatite (FAp), when hydrostatic pressure was applied from --5 to 100 kbar. }
\label{fig3}
\end{center}
\end{figure}

The NMR isotropic shielding ($\sigma_{iso}$) was computed and the results at the different pressures are reported in Figure~\ref{fig3}. The shielding on the calcium atoms, in both sites, and on the fluorine atoms was found to decrease linearly with increasing pressure (linear correlation coefficient $R^2>0.994$). The applied pressure caused a contraction of the Ca(2)--F bonds from 2.33~\AA~to 2.23~\AA~(4.3\%) (linear $R^2=0.991$). Whereas the change in bond length between Ca(2) and O(3), the next calcium nearest neighbors, was from 2.36~\AA~to 2.27~\AA~(3.8\%) (linear $R^2=0.988$). A similar trend  of decreasing shielding on Ca atoms with decreasing bond length was reported experimentally for a series of calcium containing compounds where the Ca$^{2+}$ were surrounded by oxygen atoms \cite{Dupree1997,Lin2004,Wong2006,Gervais2008}. These authors also found that the deshielding was quite linear within a class of compounds (i.e. silicates, carbonates, aluminates, etc.), with a common slope of 228 ppm/\AA. No clear correlation could instead be found by the authors when the Ca--X (X = halogen) were considered \cite{Lin2004}. A linear relationship between Ca shielding and Ca--O bond length was also observed in our calculations for FAp ($R^2>0.995$), although the calculated slopes were larger  (307 ppm/\AA~ and 414 ppm/\AA~ for Ca(1) and Ca(2), respectively). In the present work a clear linear relationship ($R^2>0.996$) was also found between the Ca shielding and the Ca--F distance, with slopes of 260 ppm/\AA~ and 352 ppm/\AA~ for Ca(1) and Ca(2), respectively.  Also, in our case, the larger slope for Ca(2) shielding vs. bond length is a further indication that the applied pressure is mainly affecting the diameter of apatite channel, which is delimited by the Ca(2) atoms.

Another point to be noted is that the FAp structure without applied pressure had a higher shielding on Ca(1) than on Ca(2), but at the highest pressure computed (100 kbar) the two shielding values were switched. This observation is somewhat in agreement with the fact mentioned by Lin \textit{et al.} (Ref.~\cite{Lin2004}) where they observed that when the electronegativity of the atom connected to the Ca increased, the shielding on the calcium decreased. Similarly here, when the distance between calcium and fluorine reached below a certain value, the very electronegative F caused the shielding on the Ca(2) to decrease faster than the one on Ca(1), for which the first nearest neighbor is instead oxygen. The value at which this ``inversion" was observed is 2.28~\AA~and it is just slightly smaller than the sum of the Ca$^{2+}$ and F$^-$ ionic radii (2.30~\AA). This also suggests that the ionic radii of O and F may play an important role in determining this observed behavior: oxygen has in fact a slightly larger ionic radius (1.38~\AA) than fluorine (1.30~\AA), so that the effects of fluorine on the shielding of Ca(2) become more relevant only at a shorter distance.\footnote{The ionic radii used are from Ref.~\cite{Shannon1976}, considering a coordination number C.N. = 3 for F, C.N. = 6 for O, and C.N. = 6 for Ca(2).}

Interestingly, while the shielding on the fluorine atoms followed a trend similar to the calcium atoms,  phosphorous displayed an opposite behavior, with its shielding increasing with increasing pressure.  As mentioned above, the change in P--O bond length was minimal in the range of pressures evaluated, thus even only a 9 ppm change in the shielding was a large change with respect to the change in bond length. This suggests that the shielding response of P is more susceptible to changes in its surrounding environment, and that this larger susceptibility together with the direct relationship between shielding and pressure is the result of the more rigid/covalent nature of the P--O bonds, with respect to the Ca--F or Ca--O. 
Finally, the shielding around the oxygen atoms was not greatly affected by pressure. This indicates a counterbalancing of the two opposite tendencies of increasing and decreasing of the shielding, when applying pressure, due to the fact that oxygens are bound to phosphorous while also bridging the calcium atoms.  
 
The pressure had also an effect on the absolute value of the shielding anisotropies, with an increasing anisotropy around Ca and F nuclei, and decreasing values for O and P. The latter finding is in agreement with the gradually decreasing difference between the P--O bond lengths of the phosphate tetrahedra, indicating a more symmetric environment around the P atoms. A decreased distortion of the phosphates from tetrahedral symmetry, with increasing pressure, was also observed experimentally by noticing a reduced site group splitting of the Raman signals of the $\nu_3$ and $\nu_4$ phosphate vibrations \cite{Williams1996}. 
The different contributions to the shielding were also evaluated and they were mainly due to paramagnetic currents tightly localized near the nuclei, except for the anions (F and O) where the induced current was more delocalized towards neighboring atoms, where partially-covalent bonds are formed. Finally, the quadrupolar parameters were only minimally affected by the pressure, with a change of less than 0.6 MHz for $C_q$ and less than 0.11 for $\eta_q$. 

\section{Conclusions}\label{conclusions}
We successfully computed the NMR chemical shifts and quadrupolar parameters of fluorapatite (FAp) and they resulted in good agreement with the available experimental data. An assignment was suggested for the $^{43}$Ca (FAp) and $^{17}$O (HAp) experimental chemical shifts (Ref~\cite{Dupree1997} and \cite{Wu1997}, respectively). The behavior of FAp under pressure and its effects on the NMR parameters were also successfully explored. The pressure caused a shrinkage of the apatite channel, with the largest changes in bond lengths observed between Ca--F and Ca--O, whereas the P--O bonds were minimally affected. A linear behavior of the NMR shielding vs. pressure was found in all nuclei, although Ca and F had an opposite slope with respect to P, while the change in the shielding on the oxygens was minimal. We believe that the opposite behavior of Ca and F when compared to P is due to the different covalency/ionicity character of the bonds involved, with the more covalent bonds having a larger shielding with decreasing atomic distance, and vice versa. Finally, our calculations can guide the calibration and the interpretation of the chemical shift of apatites in NMR experiments under high pressure.

\section{Acknowledgments}
The authors would like to acknowledge the High Performance Computer Center (HPCC) at Michigan State University (MSU) for the access to their computing resources. This work was funded by NIH grant AR056657.

\bibliographystyle{elsarticle-num}

\end{document}